\documentclass{emulateapj}
\makeatletter

\newcommand{\Rmnum}[1]{\expandafter\@slowromancap\romannumeral #1@}
\makeatother

\slugcomment{Astrophysical Journal Letters, manuscript.}

\shorttitle{The effect of PPS on the survival of exomoons}
\shortauthors{Gong, Zhou, Xie, \& Wu}

\begin{document}


\title{The effect of planet-planet scattering on the survival of exomoons
}


\author{Yan-Xiang Gong\altaffilmark{1,2}, Ji-Lin Zhou\altaffilmark{1}, Ji-Wei Xie\altaffilmark{1}, Xiao-Mei Wu\altaffilmark{2}}
\affil{$^{1}$ Department of Astronomy \& Key Laboratory of Modern Astronomy and
Astrophysics in Ministry of Education, Nanjing University,
    Nanjing, 210093, China}
\affil{$^{2}$ College of Physics and Electronic Engineering, Taishan
University, Taian, 271021, China}
\email{zhoujl@nju.edu.cn, yxgong@nju.edu.cn}



\begin{abstract}
Compared to the giant planets in the solar system, exoplanets have many
remarkable properties such as the prevalence of giant planets on eccentric
orbits and the presence of hot Jupiters. Planet-planet scattering (PPS) between
giant planets is a possible  mechanism in interpreting above and other observed
properties. If the observed giant planet architectures are indeed the outcomes
of PPS, such drastic dynamical process must affect their primordial moon
systems. In this Letter, we discuss the effect of the PPS on the survival of
their regular moons. From the viewpoint of observations, some preliminary
conclusions are drawn from the simulations. 1. PPS is a destructive process to
the moon systems, single planets on eccentric orbits are not the ideal
moon-search targets. 2. If hot Jupiters formed through PPS, their original
moons have little chance to survive. 3. Planets in multiple systems with small
eccentricities are more likely holding their primordial moons. 4. Compared to
the lower-mass planets, the massive ones in multiple systems may not be the
preferred moon-search targets if the system underwent a PPS history.

\end{abstract}


\section{Introduction}

Are there moons around exoplanets? This topic has aroused the interest of
scientists since the discovery of the extrasolar giant planets in the 1990s
\citep{williams97, barnes02, domingos06, cassidy09, donnison10, namouni10,
kaltenegger10, zhuang12}. In our solar system, all giant planets have many
moons. Natural satellites of the giant planets are divided into two categories:
regular satellites and irregular satellites. Regular satellites with little
eccentricities and orbital inclinations most probably formed in the
circumplanetary disks during the formation of the planet itself; while
irregular satellites with significant inclinations and eccentricities were most
likely to have been captured at a much later stage in the evolution of the
planet. Another difference is that regular satellites are close to their host
planets, while the irregular satellites are smaller bodies moving on distant
orbits. The prevalence of moons in the solar system implies that extrasolar
giant planets may also harbour moon systems.

Searching exomoons also have significance in exploring extrasolar lives. To
date, many extrasolar giant planets have been found in the habitable zone (HZ)
\citep{porter11, gaidos13}, but they support neither a solid nor a liquid
surface near which organisms might dwell. Exomoons of these giant planets, if
exist, may be alternative cradles for the extrasolar lives \citep{williams97,
heller12}. Based on the Kepler mission, exomoon-search methods and projects
have been proposed by several authors \citep{kipping09, simon12, kipping12,
lewis13}. Lately, several viable moon-hosting planet candidates have been
analyzed, although no compelling evidences for exomoons are found yet
\citep{kipping13}.

However, extrasolar giant planets have very different properties compared to
the giant planets in the solar system. A prominent feature is that extrasolar
giant planets have significant eccentricities, whereas four giant planets in
solar system are on nearly circular orbits. Currently, one possible  mechanism
that can explain this is planet-planet scattering (PPS) between giant planets
(\citealt{rasio96, weidenschilling96, lin97, marzari05, zhou07, ford08,
chatterjee08, juric08, raymond13}, etc.). Some of other phenomena of exoplanet
systems can also be explained by involving PPS such as the formation of hot
Jupiters (HJs) \citep{nagasawa08,nagasawa11, hirano12,beauge12}; mutual
inclinations between planets in multi-planet systems \citep{lissauer11}.

Strong evidence of PPS in multiple-planet systems such as the $\upsilon$ And
\citep{ford05} implies that PPS may be a natural dynamical process. Actually,
even in the solar system, studies by \citet{morbidelli09} have shown that PPS
between Saturn and one of the ice giants (Uranus or Neptune) need to have
occurred to reproduce the current secular properties of the giant planets,
whereas other mechanism such as smooth migration of the giant planets through a
planetesimal disk \citep{malhotra95} can not reproduce them.

If PPS indeed took place in the formation process of exoplanets, it is natural
to ask whether and how such a strong dynamical process affects the survival of
their primordial (regular) moons? In this {\em Letter}, some preliminary
studies are carried out on this topic.

\section{Initial conditions}

According to the core accretion theory, giant planets can only form outside the
snow line. However, the diversity of the protoplanetary disk, the disk-driven
migration, the presence of secular resonances such as the Lidov-Kozai effects
make the configuration of exoplanet systems diverse \citep{zhou12}. As for the
disk migration, the speed and direction depend intricately on disk physics
\citep{kley12}, so the observed orbital architectures of exoplanet systems
cannot be reproduced well even if we consider normal migration model and planet
growth model. Here, we use giant planet formation model in \citet{kokubo02} and
similar initial conditions proposed by \citet{chatterjee08}.
\begin{equation}
M_{p} = 16\pi {\left( {\frac{1}{3}\frac{{{M_{{\rm{core}}}}}}{{{M_ \odot }}}} \right)^{1/3}}
{f_g}{\Sigma _1}{a^{1/2}} + {M_{{\rm{core}}}},
\end{equation}
\begin{equation}
{a_{i + 1}} = {a_i} + K{R_{H,\;i}},
\end{equation}
where $R_{H,\;i}$ is the Hill radius of the $i$th planet and $K=4.5$,
$f_{g}=240$, $\Sigma _{1}=10$ g cm$^{-2}$, $a_{1}=3$ AU (the innermost planet
is near the snow line). We also assume that $M_{{\rm{core}}}$ satisfies a
uniform distribution between 1 and 15 ${M_ \oplus }$. The initial masses of the
planets obtained with this procedure are between about 0.25 and 1.7 $M_{J}$,
where $M_{J}$ is the mass of Jupiter. Such initial configurations can reproduce
at least two observed properties of exoplanets: 1) the observed eccentricity
distribution of exoplanets; 2) the proportion of the potential HJs ($\sim 2\%$
planets with periapse distances $<$ 0.03 AU \citep{chatterjee08}). For moons,
people are generally concerned about the Earth-like moons \citep{barnes02,
kaltenegger10, heller12}. \citet{canup06} indicates that Earth-like moons
should not form for Jupiter-mass planets. Whether it also holds in the
extrasolar systems is unclear. According to \citet{williams97}, moons
satisfying two criterions are habitable: 1) large enough ($>$ 0.12 $M_\oplus$)
to retain a substantial and long-lived atmosphere; 2) possess a strong
sheltered magnetic field. However, a remarkable discovery of the Galileo
spacecraft is that Ganymede have a magnetosphere, which suggest that some moons
($\sim 0.03\;M_\oplus$) would be immune to the atmospheric loss due to the
constant bombardment of energetic ions from its parent planet's magnetosphere.
Besides, Titan has a dense atmosphere even though its mass is 0.02 $M_\oplus$.
Based on the above analyses, we consider three kinds of moons:
$M_m=1\;M_\oplus$ (Earth-like), 0.1 $M_\oplus$ (Mars-like), 0.01 $M_\oplus$
(Moon-like).

\cite{donnison10} derived the critical semi-major axis (SMA) $a_{c}$ of a stable moon based the
Hill stability criteria, it is
\begin{eqnarray}
\frac{{{a_p}}}{{{a_{c}}}} & = & \frac{3}{x} + \frac{{{{\left( {1 + \lambda } \right)}^2}}}{\lambda }
+ \frac{{2\lambda \left( {\lambda  - 1} \right) - 3{{\left( {1 + \lambda } \right)}^3}}}{{3\lambda \left( {1 + \lambda } \right)}} \nonumber \\
& & + \frac{{\left[ {\lambda \left( {2 + 5\lambda } \right) + \frac{{12{\lambda ^2}}}{{1 + \lambda }}
+ \frac{{9\lambda \left( {1 + {\lambda ^2}} \right)}}{{{{\left( {1 + \lambda } \right)}^2}}}} \right]x}}{{3\lambda \left( {1 + \lambda } \right)}},
\end{eqnarray}
where $\lambda  = {M_m}/{M_p}$ and $x = {\left[ {\left( {{M_m} + {M_p}}
\right)/3{M_ \star}} \right]^{1/3}}$, $a_p$ is the SMA of planet, $M_{\star}$
is the mass of central star and we take $M_{\star}=M_\odot$ in this work. We
define the initial SMA of the moon as
\begin{equation}
{a_m} = f \cdot a_{c}\;\;\;\;\;(f \leq 1).
\end{equation}
The purpose of doing so is to guarantee that the escape of the moon is caused
by the scattering between giant planets rather than the planet-moon system
itself. The lower limit of SMA of the moon is the Roche radius
\citep{weidner10},
\begin{equation}
{R_{\rm roche}} = 2.44{\left( {\frac{{4\pi }}{3}} \right)^{ - 1/3}}{\left( {\frac{{{M_p}}}{{{\rho _m}}}} \right)^{1/3}},
\end{equation}
where $\rho _{m}$ is the mean densities of the moon.
$f=[0.2,\;0.4,\;0.6,\;0.8,\;1.0]$ are explored (ensure that the moon is outside
the Roche limit).

 We do a few approximations here as done by other studies on exomoons. 1. Tidal effects can be ignored if the dynamical timescale
 concerned is much shorter than the tidal evolution timescale \citep{namouni10}.
 The typical tidal evolution timescale of the moons is Gyr, which is much longer than the PPS timescales ($\sim 10^4$
 in this work). So no significant tidal evolution of the moon occurs during PPS. 2. The interactions between the moons
 (around the same giant planet) are ignored \citep{barnes02}. Though the interactions between moons may cause additional
 instabilities, other factors such as resonance (mean motion resonance, spin-orbit resonance or Laplace resonance)
 between them may guarantee stability. Above complex issues are out of the scope of this Letter. Our model contains
 3 planets, each bears an identical moon. Because irregular moons are thought most likely to have been captured at a much later
 stage in the evolution of the planet, we don't consider the effects of PPS on the survival of irregular moons. Both
 the planets and moons are on initial coplanar and circular orbits.

We use the Bulirsch-Stoer integrator in MERCURY package \citep{chambers99}, the
accuracy requirements is $10^{-12}$. For every combination of $[f,\; M_{m}]$,
we perform 100 runs,  thus a total of  1500 runs of numerical simulations are
performed in this work. The integral time is $10^6$ yr, we find it is long
enough to give the credible results. If the moon's planetocentric energy became
positive, we think it has escaped from the planet \citep{domingos06}.

\section{Simulation results and Analysis}

 Fig. 1 shows a randomly selected, representative example of dynamical evolution of planets plus moons system,
 showing both the chaotic phases and the stable final configurations. In Table 1, we show the overview of results.
 Evidently, the PPS between giant planets have violent impacts on their moons. Even if the moons are initially
 located at the inner stable region of giant planets ($f=0.2$), nearly 2/3 systems have lost their moons completely.
 For three moons of different masses, simulation results are similar, it is understandable because that the
 mass of the moon is involved in the critical SMA (Equation 3). In order to give some clues for the future
 observations, we give detailed analysis based on the architectures of extrasolar giant planet systems.
After the scattering process, the resulting systems can be classified into two
categories based on the final configuration of the giant planets: 1) two-planet
systems; 2) one-planet systems. According to above  categories, we discuss the
survival of exomoons in detail.

{\em One-planet systems.} For the total number of 500 systems with the same
mass of moons, the average fractions resulting in one-planet systems (with or
without moons) are $31.8\%$ (1 $M_\oplus$), $29.2\%$ (0.1 $M_\oplus$), $32.6\%$
(0.01 $M_\oplus$), respectively. But the fractions of remaining one planet with
a moon are only $2.2\%$, $1.2\%$ and $2.8\%$, respectively. It implies that if
a single planet on an eccentric orbit comes from a primordial multiple system
through PPS, the probability of harbouring their primordial moons is very low.

{\em Two-planet systems.} If two giant planets survived the scattering process,
they must be on well-separated orbits to ensure the long-term stability. The
total fractions of two-planet systems for three types of moons are $68.2\%$ (1
$M_\oplus$), $70.8\%$ (0.1 $M_\oplus$) and $67.4\%$ (0.01 $M_\oplus$),
respectively. In these cases, the fractions of each planet having a moon (2p +
2m) are $1.6\%$ (1 $M_\oplus$), $3.6\%$ (0.1 $M_\oplus$), $2.2\%$ (0.01
$M_\oplus$). The fractions of only one moon left (2p + 1m) are $19.2\%$ (1
$M_\oplus$), $19.4\%$ (0.1 $M_\oplus$), $16.4\%$ (0.01 $M_\oplus$). In
significant number of cases, there are only two planets left in the system (2p
+ 0m), the fractions are $47.4\%$, $47.8\%$ and $48.8\%$, respectively. By the
way, we found 8 systems where moon exchanges take place. The fractions
 of them are only $0.5\%$ in total 1500 systems. In above 8 systems, 5 of them are
 `2p + 1m' systems, 2 are `2p + 2m' systems, 1 belongs to `1p + 1m' systems.
Compared to the faithful survived moons, the promiscuous ones are
insignificant. We don't discuss the details of them in this Letter.

Evidently, the survivability of the moons in two-planet systems is higher than
that in one-planet systems. Since the `2p + 1m' systems are the dominant
outcomes where at least one stable moon survived in two-planet systems, we
focus on these cases. In Fig. 2, we give the a-e map for all the survived
planets ($M_{m}=0.1\;M_{\oplus}$). As we can see in Fig. 2, moon-bearing
planets generally have small or moderate eccentricities.

1. {\em Mass dependencies}. It can be seen from Fig. 3 that the lower mass one
of the two planets have the larger chance of bearing moons. In all the `2p +
1m' systems (275, see Table 2),  $87.6\%$ systems (241) have a moon encircling
the lower mass planet.  From the viewpoint of dynamical stability, this
conclusion is somewhat counter-intuitive because the massive planet seems
easier to hold a moon. Actually, in significant cases, the bigger planets
resulted from the merger of two planets (the total fraction is $84.4\%$ in all
the `2p + 1m' cases, see Table 2), which means that their moons have been
destroyed in the process of close encounters between two merged planets.

2. {\em Collisions vs. ejections}. Two-planet systems formed through two
channels: collisional merger of two planets or ejection of one planet. In all
the 1032 two-planet systems (including no moon systems), $\sim 80\%$ (826) are
derived from merge, $\sim 20\%$ (206) comes from ejections. One may argue:
collisional mergers are the dominant outcomes, which makes the `2p + 1m'
systems due to merges dominant. It is not the whole story. In above 206
systems, $\sim 90\%$ (185) are no-moon systems, ejection systems account for
only $7.6\%$ (21) of all the `2p + 1m' systems.\footnote{Other $8.0\%$ of `2p +
1m' systems may be unphysical--- one moon of two merged planets survived the
merge process and encircles the merged body. They are discarded in Fig.3} It
means that the survival rate of moons in two-planet systems formed due to
ejections is very low. Ejection of one planet in a system needs more frequent
close encounters than collisional merger \citep{ford08}, so the low survival
rate of the moon is understandable.

Even in all the `2p + 1m'  systems formed due to ejections, only about half of
them (12/21) have  moons encircling the bigger planets.  It shows the chaotic
nature of PPS. For example, if the smallest planet (in the initial three-planet
system) is ejected out of the system due to close encounters mainly between it
and the biggest one, their moons are destroyed,  whereas the moon of the
moderate-mass planet survived, so the moon-bearing planet is the smaller one of
the two surviving  planets.

3. {\em Inner vs. outer}. We also find that the outer planets have the larger
chance of harbouring moons than the inner ones. The ratios of outer planets
harbouring moons to the inner ones are 65/31 (1 $M_\oplus$), 56/41 (0.1
$M_\oplus$) and 55/27 (0.01 $M_\oplus$), respectively. This preference seems
independent of the initial mass distribution of three planets. Our procedure
(Equation 1) produce significant systems with $m_{1} < m_{2} < m_{3}$, where
$m_{1}$, $m_{2}$ and $m_{3}$ are the mass of initial inner, middle and outer
planet, respectively. We perform 200 additional simulations to check whether
SMA dependencies are related to the initial mass distribution of giant planets.
We randomly selected planets mass according to the observed distribution of
exoplanet masses: $dN/dM \propto {M^{ - 1.1}}$ \citep{marcy08}, the masses are
limited in the range of 0.3-1.7 $M_{J}$. We make all the systems satisfy $m_{1}
> m_{2} > m_{3}$ and $M_{m}=1\;M_{\oplus}$. Other initial conditions are similar to
Section 2. We found the ratios of outer planets harbouring moons to the inner
ones are 39/17 in all the 56 `2p + 1m' systems. It implies that collisional
mergers take place mainly in the inner region during PPS. Interestingly, within
Kepler multiple-candidate systems, the larger planets is most often the one
with the longer period for planet pairs for which one or both objects are
approximately Neptune-sized or larger \citep{ciardi13}. Besides, many planet
pairs are found near low-order mean-motion resonances \citep{lissauer11}, it
accords with resonant capture \citep{snellgrove01} of planets followed by
turbulent removal from resonance \citep{adams08}. This scenario is thought as
an origin of PPS \citep{ford08}. Consequently, SMA dependencies of exomoons in
multiple planet system can give us clues about the evolution of planet systems.

{\em HJs and giant planets in HZ.} If HJs formed by scattering mechanism
\citep{nagasawa08, nagasawa11, beauge12}, their primordial moons may be
completely destroyed, it can be seen clearly in Fig. 2. In this mechanism,
planets need to achieve great eccentricities to form a potential HJ, which
means they must undergo strong dynamical process, so the survivability of their
moons is very low. We also find that some giant planets in HZ can hold their
moons after strong PPS. In this work, the main aim is to see the impact of the
PPS on the survival rate of the moons rather than to reproduce the observed
amount of HJs or HZ giant planets. The probability of forming HJs and giant
planets in HZ are closely related to the initial locations of giant planets.
Additional simulations are performed using similar initial conditions (giant
planets) as suggested by \citet{beauge12} or \citet{raymond13}, we find similar
conclusions besides the amount of HJs and giant planets in HZ.

Finally, we found that some moons are stripped from their parent planets in the
scattering process and thus become {\em planets} encircling the star on the
stable orbits (see Fig. 2). In some cases, the remaining small planets
(original moon) and giant planets constitute a solar-like system (terrestrial
planets in inner region and gas giant planets in the outer orbits). Though the
formation of Earth-like planets through this mechanism is infrequent, it has
enlightening meanings. The origin and amount of the water on Earth is an
unresolved problem. Many attempts are made to explain the source of water
\citep{izidorro13}. If the inner planets themselves or some embryos are
originally the moons of giant planets (such as Ganymede), then significant
water content is understandable. This mechanism may be unlikely in the solar
system, but dramatic dynamical process in exoplanet systems can not exclude
this possibility.

\section{Summary}

In this Letter, we focus on the impact of PPS on the survival of exomoons.
Although there are some uncertainties in the model, some preliminary
conclusions can be drawn from simulations. 1, PPS is a destructive process to
the moon systems, planets in single planet systems, if they have large
eccentricities, are not the ideal moon-search targets. 2, If HJs formed through
PPS mechanism as suggested by many authors, their original moons have little
chance to survive. 3, Exoplanets with small eccentricities in multiple systems
more likely hold their primordial moons. 4, The massive planets in multiple
systems may not be the preferred moon-search targets if they formed by
collision-merger mechanism as suggested by \citet{lin97}. We expect
exomoon-search projects such as ``Hunt for Exomoons with Kepler'' (HEK)
\citep{kipping12} to give us interesting discoveries in the near future. The
properties of exomoons can give us clues about the evolution of planet systems
and deepen our understanding about the planet formation. Especially, exomoons
are good evidences to check the PPS hypothesis.

\acknowledgments
We thank the anonymous referee for his constructive comments
and suggestions. This work is supported by National Basic Research Program of
China (973 Program 2013CB834900), NSFC (10925313, 10833001), the Fundamental
Research Fund for Central Universities (1112020102),
 Gong Y.-X. also acknowledge the support form Shandong Provincial Natural Science Foundation,
China (No. ZR2010AQ023, ZR2010AM024).

\clearpage


\begin{table}
\begin{center}
\caption{The survival rates of moons and planets.\label{tbl-1}}
\begin{tabular}{l|l|lll|lll}
\tableline\tableline
$M_{m}$ (${M_\oplus}$)   &   $f$     & 0 moon      & 1 moon   &  2 moons  &  1 planet & 2 planets \\
\tableline
1         & 1.0     & $88\%$      & $10\%$   & $2\%$        & $24\%$   & $76\%$\\
          & 0.8     & $83\%$      & $17\%$   & $0\%$        & $31\%$   & $69\%$\\
          & 0.6     & $77\%$      & $20\%$   & $3\%$        & $30\%$   & $70\%$\\
          & 0.4     & $71\%$      & $28\%$   & $1\%$        & $35\%$   & $65\%$\\
          & 0.2     & $65\%$      & $32\%$   & $3\%$        & $39\%$   & $61\%$\\
\tableline
0.1       & 1.0     & $88\%$      & $10\%$   & $2\%$        & $31\%$   & $69\%$\\
          & 0.8     & $82\%$      & $16\%$   & $2\%$        & $26\%$   & $74\%$\\
          & 0.6     & $74\%$      & $23\%$   & $3\%$        & $25\%$   & $75\%$\\
          & 0.4     & $69\%$      & $28\%$   & $3\%$        & $31\%$   & $69\%$\\
          & 0.2     & $65\%$      & $26\%$   & $9\%$        & $33\%$   & $67\%$\\
\tableline
0.01      & 1.0     & $84\%$      & $14\%$   & $2\%$        & $39\%$   & $61\%$\\
          & 0.8     & $82\%$      & $17\%$   & $1\%$        & $28\%$   & $72\%$\\
          & 0.6     & $79\%$      & $20\%$   & $1\%$        & $34\%$   & $66\%$\\
          & 0.4     & $76\%$      & $21\%$   & $3\%$        & $35\%$   & $65\%$\\
          & 0.2     & $72\%$      & $24\%$   & $4\%$        & $27\%$   & $73\%$\\
\tableline
\end{tabular}
\tablecomments{The fraction is to 100 planetary systems. }
\end{center}
\end{table}

\begin{table}
\begin{center}
\caption{All the systems containing two giant planets and a moon.\label{tbl-1}}
\begin{tabular}{ll}
\tableline\tableline
\multicolumn{2}{c}{Total 2p + 1m (275)} \\
\tableline
                                                        &  Smaller $84.4\%$ (232) \\
                            Mergers $92.4\%$ (254)         &       \\
                                                        & Bigger $8.0\%$ (22)\\
\tableline
                                                         &   Smaller $3.3\%$ (9)\\
                           Ejections $7.6\%$ (21)          &     \\
                                                         & Bigger $4.4\%$ (12)\\

\tableline
\end{tabular}
\tablecomments{`Small' and `bigger' mean that the moon-bearing planet is the smaller one and the bigger one, respectively.}
\end{center}
\end{table}

\begin{figure}
\epsscale{0.9} \plotone{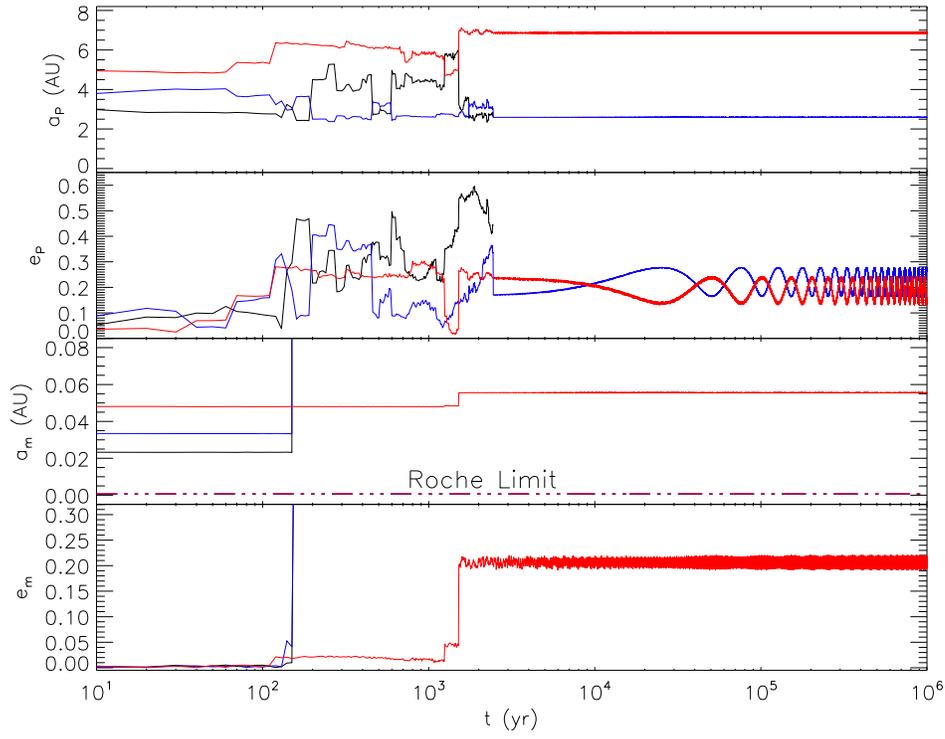} \caption{Orbital evolution of a system
composed of three giant planets and their moons (0.01 $M_{\oplus}$). The upper
two panels are semimajor axes and eccentricities of three planets. The lower
two panels are the circumplanetary semimajor axes and eccentricities of their
moons. Same color is used to denote the planet and its moon. During the close
encounters between the inner (black) and the middle (blue) planet, their moons
escaped around $167$ yrs. The inner planet was hit by the middle planet at
$\sim 2449$ yrs due to the strong perturbation of the outer one. At last, two
planets were left over in the system and the outer planet's moon survived.
\label{fig1}}\end{figure} \clearpage

\begin{figure}
\epsscale{1.0} \plotone{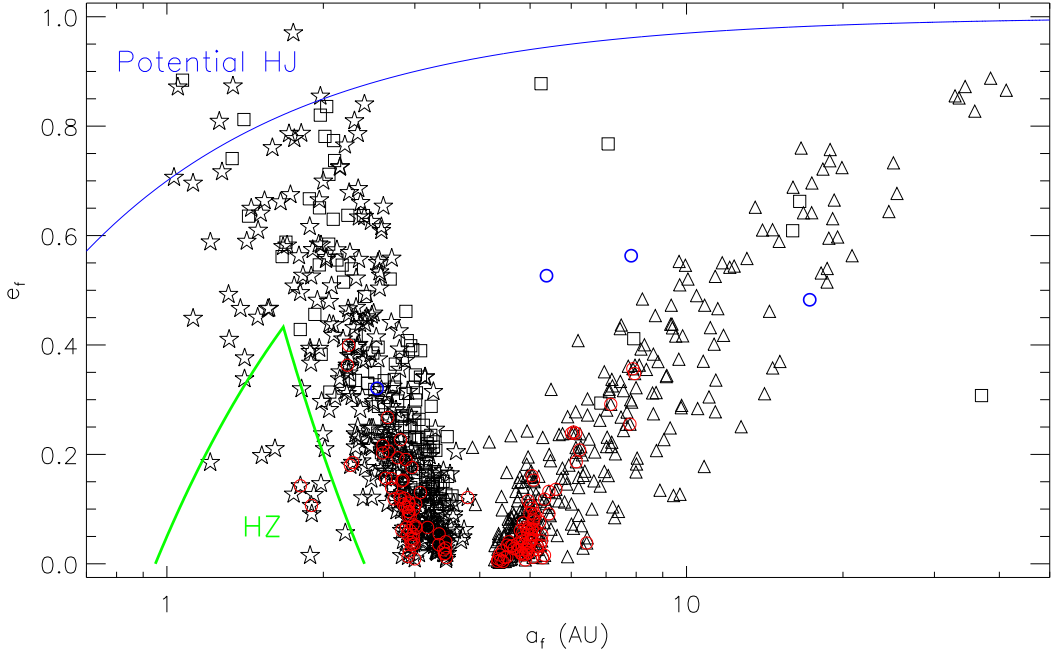} \caption{Final semimajor axis vs. eccentricity
plot for all the planets left (500 systems, $M_{moon}=0.1 M_{\oplus}$). Open
stars and triangles represent the final inner and outer planets in two-planet
systems, respectively. Open squares show the distribution of planets in
one-planet systems. Red circles are plotted on all the moon-bearing planets.
Some moons turn into planets after the scattering process, they are denoted as
open blue circles. Green lines show the habitable zone of the star
\citep{mischna00}. The planar initial configuration we adopt means that the
Lidov-Kozai mechanism don't operate, which will reduce the amount of hot
Jupiters \citep{nagasawa11}.  Here,  we define the planet with pericentric
distance $q < 0.3$ AU (blue line) as a {\em potential} hot Jupiter.
\label{fig2}}\end{figure}

\begin{figure}
\epsscale{0.6} \plotone{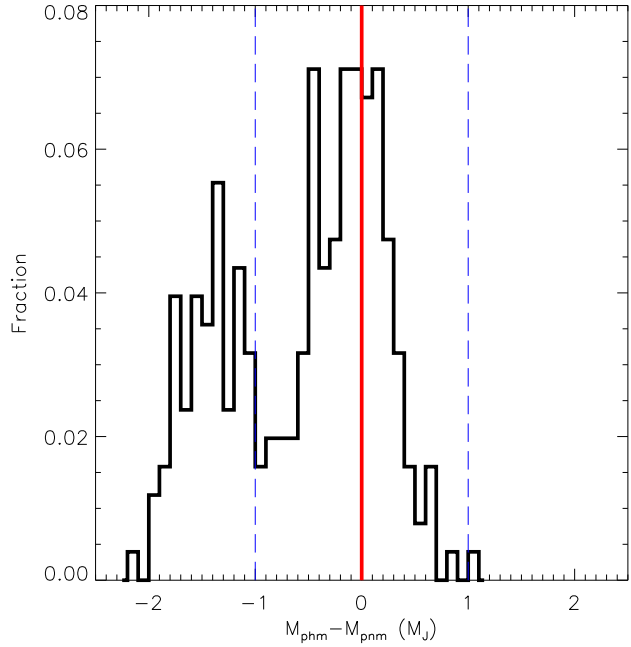} \caption{Distribution of $M_{phm}-M_{pnm}$ in
two-planet systems harbouring a moon. $M_{phm}$ denotes the mass of
moon-bearing planet and $M_{pnm}$ denotes the mass of moon-lost planet. If
$\left| {{M_{phm}} - {M_{pnm}}} \right| > 1$, the bigger one mainly comes from
the collisional merger. \label{fig3}}\end{figure} \clearpage

\end{document}